\documentclass[10pt]{article}

\usepackage{amsmath}
\usepackage{amssymb}

\usepackage{graphicx}

\usepackage{cite}

\usepackage{color} 

\usepackage[labelfont=bf,labelsep=period,justification=raggedright]{caption}

\makeatletter
\renewcommand{\@biblabel}[1]{\quad#1.}
\makeatother

\date{}

\begin{document}

\begin{flushleft}
{\Large
\textbf{An Exact Relationship Between Invasion Probability and Endemic Prevalence for Markovian SIS Dynamics on Networks}
}
\\
Robert R. Wilkinson$^{1}$, 
Kieran J. Sharkey$^{1,\ast}$, 
\\
\bf{1} Department of Mathematical Sciences, The University of Liverpool, Liverpool, UK
\\
$\ast$ E-mail: kjs@liv.ac.uk
\end{flushleft}

\section*{Abstract}

 Understanding models which represent the invasion of network-based systems by infectious agents can give important insights into many real-world situations, including the prevention and control of infectious diseases and computer viruses. Here we consider Markovian susceptible-infectious-susceptible (SIS) dynamics on finite strongly connected networks, applicable to several sexually transmitted diseases and computer viruses. In this context, a theoretical definition of endemic prevalence is easily obtained via the quasi-stationary distribution (QSD). By representing the model as a percolation process and utilising the property of duality, we also provide a theoretical definition of invasion probability. We then show that, for undirected networks, the probability of invasion from any given individual is equal to the (probabilistic) endemic prevalence, following successful invasion, at the individual (we also provide a relationship for the directed case). The total (fractional) endemic prevalence in the population is thus equal to the average invasion probability (across all individuals). Consequently, for such systems, the regions or individuals already supporting a high level of infection are likely to be the source of a successful invasion by another infectious agent. This could be used to inform targeted interventions when there is a threat from an emerging infectious disease.

\section*{Introduction}

`Compartmental' models \cite{Kermack}-\cite{Anderson1}, in which interacting individuals exist in discrete states, for example: susceptible, infectious or recovered, capture the essence of real-world host-to-host infection dynamics. Transition between states is often represented as a time-homogeneous Poisson process \cite{Bartlett}, \cite{Bailey}, which can depend on the states of other individuals. Assuming a large and evenly-mixed population, in which every individual interacts equally with every other individual, many important results have been derived by using a mean-field approximation. For example, the deterministic susceptible-infectious-recovered/removed (SIR) model exhibits an invasion threshold whereby, depending on the combined effect of the rate at which individuals make infection-causing contacts and the rate at which infected individuals recover, a small number of initial infecteds will either cause a significant outbreak, the size of which can be calculated and is often referred to as the `final size', or the infection rapidly dies out \cite{Kermack}. Likewise, the deterministic susceptible-infectious-susceptible (SIS) model also exhibits an invasion threshold such that, depending on the same factors, the infection either persists at some constant endemic level or rapidly dies out \cite{Weiss}. The way in which these thresholds, and the final size and endemic prevalence, are affected when an immunisation process is included has been investigated. By comparing these and similar models with real statistical data it has also been possible to quantify the invasion threshold, and expected impact on the population, for various diseases \cite{Hethcote1}, \cite{Anderson2}. Therefore, we have some way of determining optimum vaccination policies for the eradication or control of specific diseases.  

To reflect the probabilistic nature of the real-world process of invasion, and disease transmission in general, stochastic descriptions are required. Moreover, the probability of invasion from a single infectious individual clearly depends on that individual's particular relationships with others, e.g. some individuals may be better connected than others. In order to capture such heterogeneity, the population can be represented as a contact network \cite{Newman} which defines, for each individual, the subset of the population with which it has direct contact. If the population is then assumed to be infinite such that the number of neighbours with which an individual has contact is described by a probability distribution, it is sometimes possible to compute the invasion probability and (fractional) final size for the stochastic SIR model by utilising percolation theory \cite{Newman}-\cite{Kenah}. For finite populations, there are many numerical methods by which to measure the final size distribution~\cite{House}. However, finite contact networks do pose conceptual difficulties; indeed, endemic behaviour is often associated with non-trivial stationary distributions which for many finite systems do not exist, while the theoretical definition of invasion probability depends on a framework which posits an infinite population such that invasion corresponds to indefinite persistence. 

In the next section we will introduce the network-based SIS stochastic model and explain, following Harris \cite{Harris2}, how it can be represented graphically. We will also show how such a graphical representation can be used to prove an important equation, which we state, and which expresses the property known as `duality'. This property was discovered by Holley and Liggett \cite{Holley}, and Harris \cite{Harris3}, but it is usually discussed in relation to undirected non-weighted networks. We find that it is also relevant in the context of directed weighted networks. In `Theoretical Results' we will define and justify, for an arbitrary strongly connected network, exact mathematical quantifiers for both endemic prevalence and invasion probability. We also prove an exact relationship between them (see Pastor-Satorras and Vespignani~\cite{Pastor1} for a discussion of endemicity in random scale-free networks, and Gilligan and van den Bosch~\cite{Gilligan} for an overview of `invasion and persistence' in epidemiological models). In `Numerical Simulation' we will discuss how these quantifiers are to be measured through stochastic simulation and provide some examples. Notably, we will illustrate our theoretical results on the largest strongly connected (5,119 node) component $T_{\small{\mbox{ex}}}$ (network data S1) of a particular heterogeneous transmission network. The full network comprises 11,480 nodes and was generated from simulations on a complex model of the spread of H5N1 avian influenza through the British poultry flock  \cite{Sharkey},  \cite{Jonkers}. It exhibits extensive heterogeneity including complex spatial structures, heterogeneous transmission strengths varying over many orders of magnitude, clustering and directed links. It therefore serves as a somewhat rare example of `realistic' epidemic contact structures.

\section*{Markovian SIS Dynamics on a Contact Network}

 In Markovian SIS dynamics, an individual is able to flip repeatedly between two states: susceptible and infectious.  This happens via a locally influenced time-homogeneous Poisson transmission process and an individual-specific time-homogeneous Poisson recovery process (on recovery an individual returns to the susceptible state).  In the context of individuals interacting in this way on a regular square lattice the SIS model is also known as the contact process \cite{Harris1} (see Liggett  \cite{Liggett} for theoretical results, and Durrett and Levin  \cite{Durrett1} for an ecological perspective). We will consider the dynamics of the Markovian SIS model on the full set of networks which are finite and static (our theoretical results will apply to the subset that are strongly connected). A generic weighted network (that can also be directed) will be denoted by a matrix $T$, where $T_{ij}$ indicates the rate parameter of the Poisson process in which individual $j$ infects individual $i$, assuming $j$ is infectious and $i$ is susceptible ($i,j \in \{1,2,\ldots,N\}$ where $N$ is the population size). Thus, $T$ combines the rates at which the individuals interact with the probability that infection occurs when an infectious individual contacts a susceptible individual. In this way, $T$ captures features of the network and the infectious agent. We will also refer to a vector $\Gamma=(\gamma_1,\gamma_2,\ldots,\gamma_N)$ where $\gamma_i$ is the rate parameter of the Poisson process in which $i$ recovers when it is infectious.

Assuming the system is in a specific stochastic state, let $\lambda_i$ be the infectious pressure on individual $i$ such that $i$ is on course to be infected via the `first arrival' of a Poisson process with rate parameter $\lambda_i$. Similarly, let $\mu_i$ be the recovery pressure on individual $i$ such that $i$ is on course to recover via the `first arrival' of a Poisson process with rate parameter $\mu_i$. We can now define the stochastic model with the following equations:
\begin{eqnarray}  \nonumber
\lambda_i&=&\sum_{j=1}^NT_{ij}I_jS_i \\
\mu_i&=&\gamma_iI_i
\end{eqnarray}
where $I_i=1$ if $i$ is infectious and $S_i$=1 if $i$ is susceptible, and both are zero otherwise. Given an initial configuration, such that the state of each individual is known, direct simulation can be employed to produce statistically accurate realisations of the stochastic model for any $T$ and $\Gamma$.

This framework, which represents the dynamics of several sexually transmitted diseases  \cite{Hethcote2}-\cite{Keeling1} and computer viruses  \cite{Kephart1}-\cite{Balthrop}, exhibits the phenomena of both invasion and endemic prevalence (See Durrett  \cite{Durrett2} for a discussion of methods and results relating to SIS dynamics on random scale-free networks). 

\subsection*{Graphical Representation}

\begin{figure}
 \centerline{\includegraphics[width=0.4\textwidth]{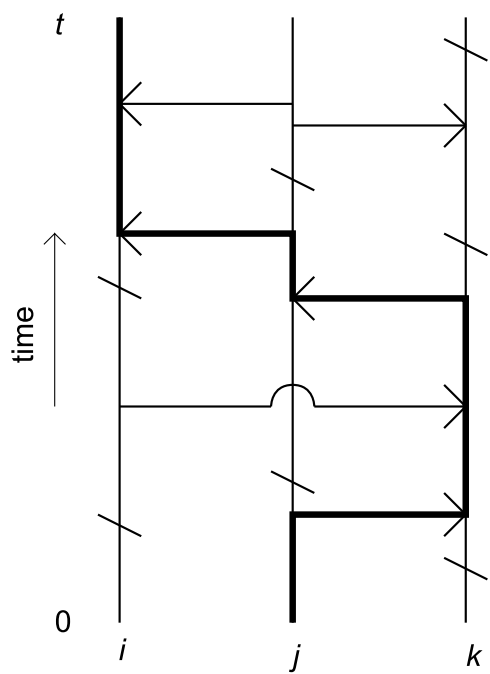}}
\caption{{\bf  A realisation of the graphical representation of Markovian SIS dynamics on a fully connected network of three individuals, $\boldsymbol{i}$, $\boldsymbol{j}$ and $\boldsymbol{k}$ (up to time $\boldsymbol{t}$).} The vertical lines are the time lines corresponding to each individual. The short diagonal lines indicate the points of cure and the horizontal arrows are the arrows of infection. A path from 0 on $j$'s time line to $t$ on $i$'s time line is shown in bold.}
\label{F1}
\end{figure}

Harris \cite{Harris2} showed that the contact process, and equivalently Markovian network-based SIS dynamics, can be fruitfully represented as follows: Each member of the population is assigned its own positive real number line or `time line'. Next, a Poisson point process with intensity $\gamma_i$ is placed on the time line corresponding to each individual $i\in V$ where $V =  \{1,2,\ldots,N\} $. Following Grimmett \cite{Grimmett}, such points are called `points of cure'. Then, for each ordered pair of individuals $(i,j)$, such that $T_{ij} \neq 0$, we place arrows going from $j$'s time line to $i$'s time line according to a Poisson point process with intensity $T_{ij}$. These arrows are called `arrows of infection'. Under such a representation, the probability that there is at least one path from 0 on a time line corresponding to an individual in subset $A\subseteq V$ to $t$ on the time line corresponding to individual $k$ is equal to the probability that $k$ will be infected at time $t$ when only the members of $A$ are initially infected (paths take into account the direction of time and the directions of the arrows of infection and are blocked by the points of cure). An example of this kind of representation, for a fully connected network of three individuals, is given in figure~\ref{F1}.

\subsection*{Notation}

Assume that we have a network $T$, with an associated vector of recovery parameters $\Gamma$. Now, graphically representing Markovian SIS dynamics on this network, and adopting the notation of Harris \cite{Harris1}, let $\xi_t^A$ (with $A \subseteq V$) be the set of individuals such that $i \in \xi_t^A$ if and only if there is at least one path from 0 on a time line corresponding to an individual in $A$ to $t$ on $i$'s time line (see figure~\ref{F1} for an example of a path). We will use $P_{T,\Gamma}$ to denote the appropriate probability measure. Thus, the probability that at least one member of a subset $B$ will be infected at time $t$, given that only the members of subset $A$ are initially infected, i.e. at $t=0$, can be expressed as $P_{T,\Gamma}(\xi_t^A \cap B \neq \emptyset)$.

\subsection*{The Duality Property for General Networks}

The property of `duality' pertaining to Markovian SIS dynamics (and the contact process) can be expressed as follows:
\begin{equation}
\label{dual}
P_{T,\Gamma}(\xi_t^A \cap B \neq \emptyset)= P_{T^T,\Gamma}(\xi_t^B \cap A \neq \emptyset)
\end{equation}
where $A,B \subseteq V$, $T$ is an arbitrary weighted contact network and $T^T$ is the transpose of $T$. Note that for undirected networks $T=T^T$ which simplifies the relationship and, in this case, the process is said to be `self-dual'.

To understand equation~\ref{dual}, it is simplest to firstly consider the case where $T$ is undirected and where the subsets contain a single individual; $A=j$ and $B=i$. Then the probability of existence of a path from $j$ at time 0 to $i$ at time $t$ is expressed by the left hand side of equation~\ref{dual}. With reference to figure~\ref{F1}, reversing the arrow of time (turning the diagram upside down), we see that the probability that a path exists from individual $i$ at time 0 to $j$ at time $t$ is the same calculation since the Poisson point processes are the same in reverse. This is expressed by equation~\ref{dual}. 

When the network $T$ possesses an asymmetry, the reverse process needs to be computed on the transposed network to be equivalent since the direction of infection is reversed. More generally, the probability that there exists at least one path from an individual in set $A$ at time 0 to an individual in set $B$ at time $t$ is equal to the probability that there is at least one path from an individual in set $B$ at time 0 to an individual in set $A$ at time $t$ in the transposed network, and this statement is represented by equation~\ref{dual}. 

\section*{Theoretical Results}

\subsection*{The Quasi-Stationary Distribution as a Quantifier of Endemic Prevalence}

\begin{figure}
 \centerline{\includegraphics[width=1.0\textwidth]{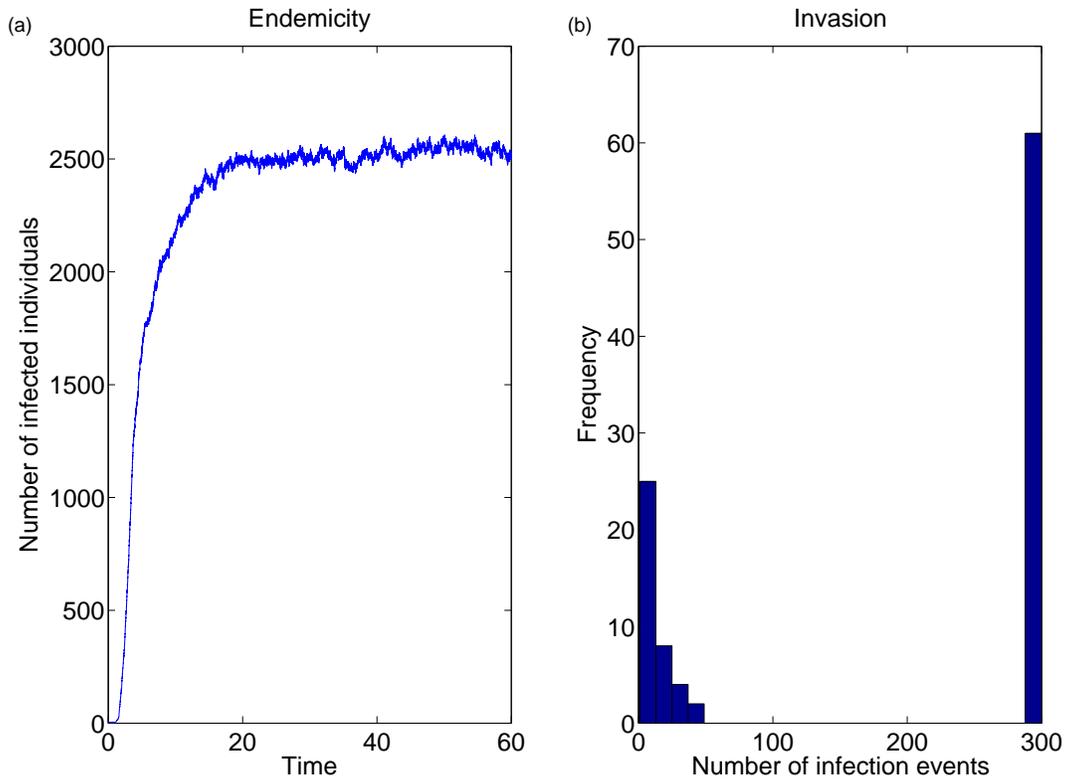}}
\caption{{\bf Numerical data from simulations of Markovian SIS dynamics on our example network $\boldsymbol{T_{\mbox{\small{ex}}}}$ (network data S1)}. (a) is a plot of the total number of infected individuals against time in a simulation where the outbreak was initiated on a single infectious individual. (b) is a histogram of the number of infection events in 100 simulations of an outbreak, which were allowed to run up to a maximum of 300 infection events, initiated on the same individual each time. In both cases, the weighted network matrix was multiplied by 0.01 and the recovery rate was set to unity for all individuals.}
\label{F2}
\end{figure}

An immediate problem with finite systems is that there are no genuine stationary distributions corresponding to endemic infection because the long-term behaviour is always guaranteed extinction (of the infectious agent). However, from a practical point of view, it is sometimes possible to obtain something like the endemic stationary distribution; figure~\ref{F2}a illustrates this clearly for our example network $T_{\small{\mbox{ex}}}$ (network data S1) (figure~\ref{F2}b illustrates the probabilistic nature of the invasion process which is discussed in subsequent sections). While a genuine stationary distribution does not exist, it is clear that we can still measure something similar to endemic prevalence through stochastic simulation since ultimate extinction is, in this case, very unlikely on short time scales.

From a theoretical perspective, the situation can be made precise. Our system is finite and Markovian with a single absorbing state (extinction of the infection). Also, if the network under consideration is strongly connected such that infection can be transmitted, via some route, from any individual to any other individual, then the transient states form a single commuting class. In this case, if we initiate the system in a transient state and condition on the survival of the infection, then the system tends to a unique distribution referred to as the quasi-stationary distribution (QSD)  \cite{Darroch1}-\cite{Nasell1}. Let us denote by $A$ an arbitrary subset of the population represented by a strongly connected network $T$, and let $\Gamma$ be the vector of individual-specific recovery parameters. We can now unambiguously define $P_{T,\Gamma}^A(\mbox{quasi-prevalence})$ to be the probability that at least one individual in subset $A$ of the network is infectious in the QSD. In the notation of the graphical representation introduced earlier, we can write:

\begin{eqnarray} \nonumber
P_{T,\Gamma}^A(\mbox{quasi-prevalence})&=& \lim \limits_{t \to \infty}⁡ P_{T,\Gamma}(\xi_t^V \cap A \neq \emptyset \mid \xi_t^V \cap V \neq \emptyset) \\
&=&  \lim \limits_{t \to \infty}⁡\frac{ P_{T,\Gamma}(\xi_t^V \cap A \neq \emptyset)}{ P_{T,\Gamma}( \xi_t^V \cap V \neq \emptyset)}  
\end{eqnarray}
since $\xi_t^V \cap A \neq \emptyset \Rightarrow \xi_t^V \cap V \neq \emptyset$.

It can be argued that the quasi-stationary distribution has practical relevance (i.e. is a good `representation' of the endemic situation \cite{Nasell1}) if the rate of convergence to this distribution, when conditioning on non-absorption, is rapid compared to the rate at which the system decays to inevitable absorption when it is `initiated' in the QSD \cite{Darroch1}. This can often be the case for SIS dynamics for which, according to N\r{a}sell \cite{Nasell1}, `it is easy to find examples where the expected time to extinction even for a rather small population exceeds the age of the universe'. More specifically, Simonis \cite{Simonis} has shown that the contact process on large but finite multi-dimensional homogeneous square lattices, where the initial state is all-infected, will be near to the upper invariant measure of the corresponding infinite process, restricted to the finite set, for most of its lifetime (assuming the corresponding infinite process is supercritical).

 Let us consider the following quantities for an arbitrary strongly connected network $T$ and arbitrary $\Gamma$:

\begin{enumerate}
\item $ P_{T,\Gamma}(\xi_t^V \cap A \neq \emptyset)=$ The probability that at least one member of subset $A$ is infected at time $t$ given that all individuals are infected at $t=0$.
\item $ P_{T,\Gamma}(\xi_t^V \cap V \neq \emptyset)=$ The probability that the infection survives to time $t$ given that all individuals are infected at $t=0$.
\item $ P_{T,\Gamma}(\xi_t^V \cap A \neq \emptyset)/ P_{T,\Gamma}( \xi_t^V \cap V \neq \emptyset)=$ The probability that at least one member of subset $A$ is infected at time $t$ given that all individuals are infected at $t=0$ and given that the infection survives to time $t$. 
\end{enumerate}
Note that in the limit as $t \to \infty$ quantity 3 is equal to $P_{T,\Gamma}^A(\mbox{quasi-prevalence})$, the probability that at least one member of $A$ is infected in the QSD.

\begin{figure}
 \centerline{\includegraphics[width=0.9\textwidth]{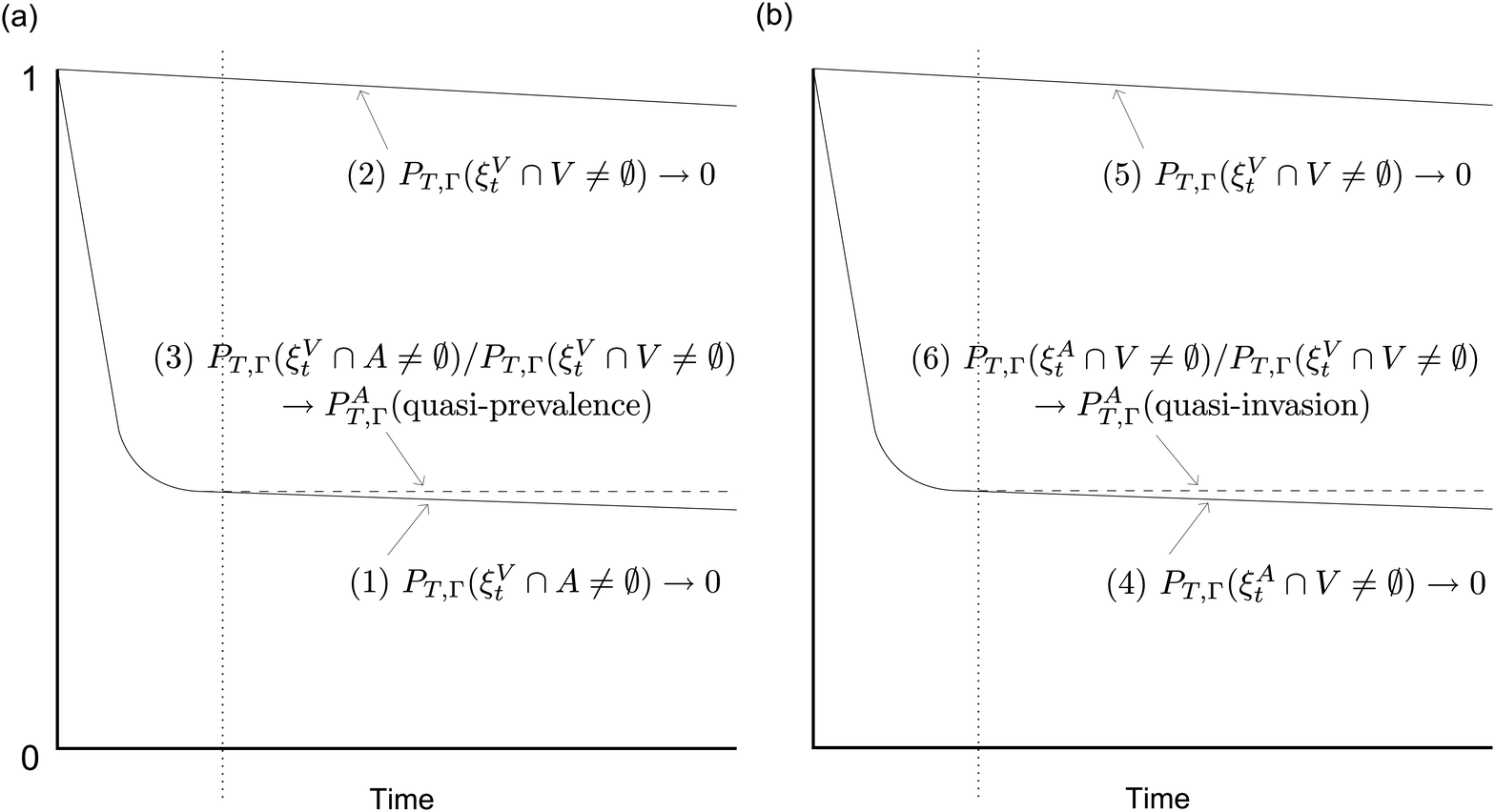}}
\caption{{\bf Here we illustrate how it is possible for the quantifiers $\boldsymbol{P_{T,\Gamma}^A(\mbox{quasi-prevalence})}$ and $\boldsymbol{P_{T,\Gamma}^A(\mbox{quasi-invasion})}$ to capture critical information about the model.} If $T$ is undirected then these quantifiers are numerically the same and have equal practical relevance (as is seen by assuming that $T$ is the same undirected network in (a) and (b), above).}
\label{F3}
\end{figure}

 In figure~\ref{F3}a, the way in which these three quantities may vary with respect to time is illustrated for a scenario in which the QSD has practical relevance. In such a scenario, the quantifier $P_{T,\Gamma}^A(\mbox{quasi-prevalence})$ is able to capture the value at which $P_{T,\Gamma}(\xi_t^V \cap A \neq \emptyset)$ initially `plateaus' before its slow decay to zero.

We have defined $P_{T,\Gamma}^A(\mbox{quasi-prevalence})$ in terms of the process which occurs when all individuals are initially infected. We could, however, define $P_{T,\Gamma}^A(\mbox{\mbox{quasi-prevalence}})$ in terms of the process which occurs when only the members of $B \subset V$ are initially infected since the QSD, as defined by Daroch and Seneta \cite{Darroch1}, \cite{Darroch2}, is independent of initial conditions. Nonetheless, the process which occurs when all individuals are initially infected is unique in that it has the maximum expected time to extinction across all possible initial states (this follows from $\xi_t^B \subseteq \xi_t^V$ (with $B \subset V)$ - see, for example, Grimmett \cite{Grimmett}). Therefore, this is the non-conditioned process which is most appropriately described by the QSD.

\subsection*{The Equivalent Quantifier of Invasion Probability}

In the context of finite contact networks, invasion probability has not been given a rigorous theoretical definition (see N\r{a}sell  \cite{Nasell2} for a discussion of the threshold phenomenon in the stochastic SIS model). For dynamics with a genuine stationary endemic distribution, which (in this context) can only exist when the population size is infinite, the probability of invasion can be defined as the probability that the infection will persist indefinitely. For finite populations, this is complicated because we have to distinguish between infections that fail to invade and infections which successfully invade but then subsequently die out.

In this section we show that the quantifier of invasion probability, which is as equally meaningful and relevant as our quantifier for endemic quasi-prevalence, for outbreaks initiated on the members of a subset $A$ of a strongly connected network $T$ (with associated vector $\Gamma$) can be defined as:
\begin{equation}
P_{T,\Gamma}^A (\mbox{quasi-invasion})= \lim \limits_{t \to \infty}⁡\frac{ P_{T,\Gamma}(\xi_t^A \cap V \neq \emptyset)}{ P_{T,\Gamma}( \xi_t^V \cap V \neq \emptyset)} .   
\end{equation}
Note that the definition of $P_{T,\Gamma}^A(\mbox{quasi-invasion})$ implies that quasi-invasion is certain when all individuals are initially infected. Our definition corresponds to the intuitive notion of invasion, i.e. a `large' outbreak, or the `attainment' of the QSD.

 Let us now consider the following quantities:
\begin{enumerate}
\item[4.] $ P_{T,\Gamma}(\xi_t^A \cap V \neq \emptyset)=$ The probability that the infection survives to time $t$ given that only the members of $A$ are infected at $t=0$.
\item[5.] $ P_{T,\Gamma}(\xi_t^V \cap V \neq \emptyset)=$ The probability that the infection survives to time $t$ given that all individuals are infected at $t=0$.
\item[6.] $ P_{T,\Gamma}(\xi_t^A \cap V \neq \emptyset)/ P_{T,\Gamma}( \xi_t^V \cap V \neq \emptyset)=$ The quotient of quantities 4 and 5.
\end{enumerate}
It follows from duality that the three quantities, 4, 5 and 6, are all equal respectively to the three quantities, 1, 2 and 3, provided that we transpose $T$. Note also that, in the limit as $t \to \infty$, quantity 6 is equal to $P_{T,\Gamma}^A(\mbox{quasi-invasion})$.

Quantity 4 denotes the survival probability up to time $t$. We see in figure~\ref{F3}b that this `plateaus' in exactly the same way as quantity 1 for the transposed network (figure~\ref{F3}a). This plateau, which is captured by quantity 6 in the limit as $t \to \infty$, i.e. $P_{T,\Gamma}^A(\mbox{quasi-invasion})$, corresponds to the `achievement' of the QSD and thus with successful quasi-invasion. 

Our quantifier of invasion probability can be generalised as:
\begin{equation}
P^X (\mbox{quasi-invasion})=  \lim \limits_{t \to \infty}⁡\frac{P_S^X(t)}{P_S^{\small{\mbox{max}}} (t)}
\end{equation}
where $X$ is a transient stochastic state in which the infection is present.  $P_S^X (t)$ is the probability of survival to time $t$ given that the system is initiated in state $X$, and $P_S^{\small{\mbox{max}}} (t)$ is the probability of survival to time $t$ given that the initial state is that which maximises the expected time to extinction. In this form, the definition becomes applicable to other Markovian infection dynamics (on strongly connected networks) which permit endemic behaviour, e.g. susceptible-infected-removed-susceptible (SIRS) dynamics. It is the existence of a unique QSD which enables our definition to capture the probability of invasion in the same way as for SIS dynamics. Note that the definition of quasi-prevalence can also be generalised to any infection model with a unique QSD.

\subsection*{The Prevalence-Invasion Relationship}

Our main result can be stated as the following (prevalence-invasion) relationship:  
\begin{equation}
\label{previnv}
 P_{T,\Gamma}^A(\mbox{quasi-invasion})=P_{T^T,\Gamma}^A(\mbox{quasi-prevalence}) 
\end{equation}
   for any subset $A$ of a weighted and strongly connected contact network $T$, conditional on Markovian SIS dynamics. 

Equation~\ref{previnv} can be re-written as:
\begin{equation}
 \lim \limits_{t \to \infty}⁡\frac{ P_{T,\Gamma}(\xi_t^A \cap V \neq \emptyset)}{ P_{T,\Gamma}( \xi_t^V \cap V \neq \emptyset)}    =\lim \limits_{t \to \infty}⁡\frac{ P_{T^T,\Gamma}(\xi_t^V \cap A \neq \emptyset)}{ P_{T^T,\Gamma}( \xi_t^V \cap V \neq \emptyset)}    \end{equation}
which holds because of the property of duality. 

Note that for a single individual we have:
\begin{equation}
 P_{T,\Gamma}^i(\mbox{quasi-invasion})=P_{T^T,\Gamma}^i(\mbox{quasi-prevalence}),
 \end{equation}
that is, the probability of quasi-invasion from a given individual $i$ is equal to the probability that it is infected in the QSD (in the transposed network). By summing over all $i \in V$ and dividing by $N$ we get
\begin{equation}
 P_{T,\Gamma}^{\small{\mbox{global}}}(\mbox{quasi-invasion})=P_{T^T,\Gamma}^{\small{\mbox{global}}}(\mbox{quasi-prevalence})
 \end{equation}
where $P_{T,\Gamma}^{\small{\mbox{global}}}(\mbox{quasi-invasion})$ is the probability of quasi-invasion given that the infection is seeded on a single individual selected uniformly at random, and $P_{T,\Gamma}^{\small{\mbox{global}}}(\mbox{quasi-prevalence})$ is the average fraction of the population that are infected in the QSD. An implication of the global-level relationship is that, for (strongly connected) directed networks, reversing the transmission processes (transposing $T$) will result in an interchange between these two quantifiers without affecting the `stability' of the quasi-stationary behaviour, i.e. the rate at which the system decays to inevitable extinction when `initiated' in the QSD is the same for $T$ and its transpose. This can be understood by observing that $P_{T,\Gamma}(\xi_t^V \cap V \neq \emptyset)=P_{T^T,\Gamma}(\xi_t^V \cap V \neq \emptyset) \quad \forall t \ge 0$. Also note that $P_{T,\Gamma}(\xi_t^i \cap i \neq \emptyset)=P_{T^T,\Gamma}(\xi_t^i \cap i \neq \emptyset) \quad \forall t \ge 0$, i.e. given the infection is initiated by individual $i$, the probability that $i$ is infectious at any $t \ge 0$ is the same for $T$ and its transpose.

For the case where $T$ is undirected ($T=T^T$), the relationship implies that:
\begin{equation}
P_{T,\Gamma}^A(\mbox{quasi-invasion})=P_{T,\Gamma}^A(\mbox{quasi-prevalence}),
\end{equation}
and for a single individual $i \in V$:
\begin{equation}
P_{T,\Gamma}^i(\mbox{quasi-invasion})=P_{T,\Gamma}^i(\mbox{quasi-prevalence}),
\end{equation}
and globally:
\begin{equation}
 P_{T,\Gamma}^{\small{\mbox{global}}}(\mbox{quasi-invasion})=P_{T,\Gamma}^{\small{\mbox{global}}}(\mbox{quasi-prevalence}).
\end{equation}

\section*{Numerical Simulation}

\subsection*{Measurement $P_{T,\Gamma}^i(\mbox{quasi-prevalence})$ by Stochastic Simulation}

The  probability $P_{T,\Gamma}^i (\mbox{quasi-prevalence})$ that individual $i$ is infectious in the QSD is equal to the proportion of time for which $i$ is infected after the system has `reached' this distribution, assuming the infection does not die out. Therefore, we measure this as
\begin{equation}   P_{T,\Gamma}^i (\mbox{quasi-prevalence}) \approx\frac{1}{\tau}  \sum_{k=1}^n \Delta t_k I_i^k \end{equation}
where $n$ is a large number of simulated consecutive global events which occur when the probabilities for the system states obey the QSD. $\Delta t_k$ is the simulated time between the $(k-1)^{th}$ event and the $k^{th}$ event, and $I_i^k=1$ if $i$ is in the infectious state between these events but is zero otherwise. The total simulated time is denoted by $\tau=\sum_{k=1}^n \Delta t_k$. Obviously, the larger we can make $n$, the more exact our measurement becomes. Notice that the events corresponding to a change in the state of individual $i$, and thus to a change in the value of $I_i^k$, represent a very small fraction of the total global simulated events (assuming the population size is not extremely small).

 In practice, we run the simulation for a sufficiently long time such that the system is likely to be described by the QSD before we start to compute  $P_{T,\Gamma}^i (\mbox{quasi-prevalence})$. For example, were our simulation to produce an infectious time line similar to the one in figure~\ref{F2}a we would only include the data from after, say, $t=30$ in our computation. Note that in every simulation there is always the possibility of extinction, even after the QSD has been `reached'. However, we can safely discard any post-extinction data since the quantity we are interested in is conditioned on non-extinction. This method of measurement is only valid for systems in which quasi-stationary behaviour can be easily identified i.e. systems for which the QSD has significant practical relevance.

\subsection*{Measurement of $P_{T,\Gamma}^i(\mbox{quasi-invasion})$ by Stochastic Simulation}

In measuring the probability of quasi-invasion through stochastic simulation, the key requirement is separating major outbreaks from minor outbreaks. Therefore, we look for dichotomised behaviour in relation to the time until extinction by carrying out large numbers of simulations, each initiated in the same stochastic state. For example, we can measure $P_{T,\Gamma}^i(\mbox{quasi-invasion})$ as the fraction of simulations in which an outbreak initiated by individual $i$ persists for some sufficiently large number $E_I$ of infection events, such that a clear distinction can be made between the simulations in which long-term persistence occurs and the ones in which it does not. This is because $P_{T,\Gamma}^i(\mbox{quasi-invasion})$ corresponds to the value at which $P_{T,\Gamma}(\xi_t^i \cap V \neq \emptyset)$ initially plateaus. Without observation of this dichotomised behaviour it may not be possible to classify every simulation. For example, the histogram of infection events per simulation in figure~\ref{F2}b enables us to clearly identify the simulations which exhibit long-term persistence, even though we have only tracked each simulation to the 300th infection event i.e. $E_I=300$. If such a histogram cannot be produced from the simulations, $E_I$ can be increased in order to try and `uncover' the dichotomised behaviour.

In general, so long as this kind of dichotomised behaviour is found, the practical issues which emerge in measuring quasi-invasion probability by stochastic simulation for the SIS framework are minor. In other words, this method of measurement is valid, and easily carried out, in systems where $P_{T,\Gamma}(\xi_t^i \cap V \neq \emptyset)$ significantly plateaus i.e. systems where $P_{T,\Gamma}^i(\mbox{quasi-invasion})$ has significant practical relevance.

\subsection*{Simulations on our Example Network}

\begin{figure}
 \centerline{\includegraphics[width=1.1\textwidth]{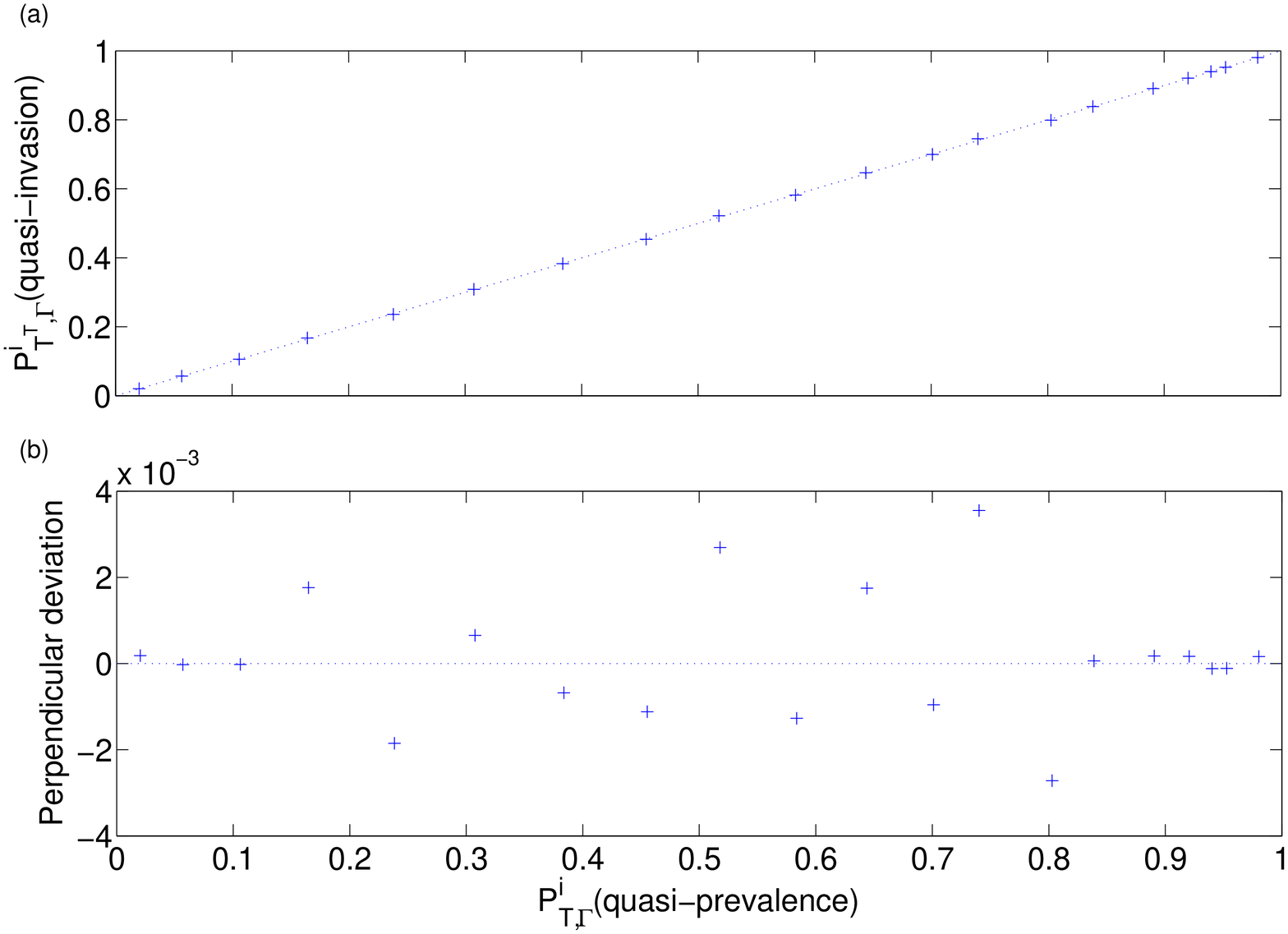}}
\caption{{\bf  Measurements of $\boldsymbol{P_{T^T,\Gamma}^i(\mbox{quasi-invasion})}$ and $\boldsymbol{P_{T,\Gamma}^i(\mbox{quasi-prevalence})}$ for a single individual ($\boldsymbol{i}$ = node 2332) in our example network $\boldsymbol{T_{\mbox{\small{ex}}}}$.} The recovery rate was set to unity for all individuals while the multiplier of the network matrix was varied. In (a), these two quantifiers are plotted against each other for each of 20 different multipliers of the network matrix. The faint dashed line indicates equality. On this scale it is not possible to determine any deviation from the equality of the two quantities. (b) is a `zoomed-in' view of the perpendicular deviation of each of the data points from the straight line (equality), in the bottom right to top left direction.}
\label{F4}
\end{figure}

By varying a scalar multiplier of a network matrix we can attempt to investigate infections of varying transmissibility spreading through the same population. Figure~\ref{F4} illustrates our theoretical results, via this method of investigation, for a single individual in our example network $T_{\small{\mbox{ex}}}$ (network data S1), clearly showing the relationship between quasi-invasion probability and endemic quasi-prevalence. We supply these numerical results to illustrate the prevalence-invasion relationship. We do of course obtain the same qualitative behaviour for any strongly connected network as proved in `Theoretical Results'.

To obtain measurements of $P_{T,\Gamma}^i(\mbox{quasi-prevalence})$ for individual $i=2332$, 100 simulations were run for each of 20 different multipliers of $T_{\mbox{\small{ex}}}$, and in each simulation the first 10 million events, out of 11 million, were discarded. Each simulation was initiated with sufficient infected individuals such that the probability of early extinction was negligible. As extinction never occurred, none of the data was discarded. 

To obtain measurements of $P_{T^T,\Gamma}^i(\mbox{quasi-invasion})$, 1 million simulations were run for each of the same 20 multipliers. In each simulation, the infection was initiated on individual $i=2332$ and the process was tracked for a maximum of 500 infection events.

\subsection*{Simulations on a Small Square Lattice}

\begin{figure}
 \centerline{\includegraphics[width=1.2\textwidth]{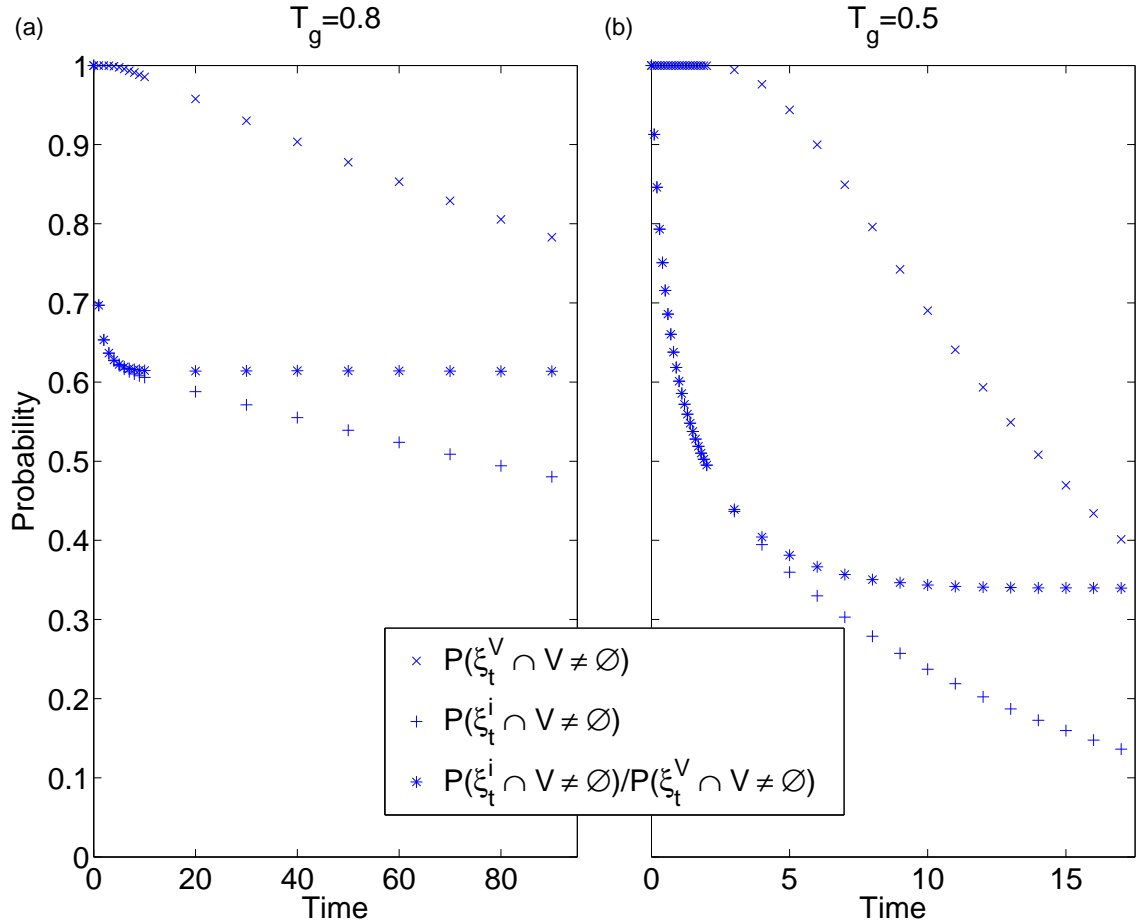}}
\caption{{\bf Here we illustrate a method of measurement, through stochastic simulation, for $\boldsymbol{P_{T,\Gamma}^i(\mbox{quasi-invasion)}}$ ($\boldsymbol{=P_{T,\Gamma}^i(\mbox{quasi-prevalence})}$) where $\boldsymbol{T}$ is an undirected and homogeneously weighted square lattice of 25 individuals (we look for the value towards which $\boldsymbol{P_{T,\Gamma}(\xi_t^i \cap V \neq \emptyset)/ P_{T,\Gamma}( \xi_t^V \cap V \neq \emptyset)}$ converges).} For (a), the global transmission parameter ($T_g$) was set to 0.8. For (b), the global transmission parameter was 0.5. The global recovery parameter was set to 1 in both cases. The figure illustrates how these quantifiers are well defined but not always practically relevant.}
\label{F5}
\end{figure}

An undirected and homogeneously weighted square lattice $T$ of 25 individuals was investigated (see figure~\ref{F5}). Due to the small population size, the probability of extinction on a relatively short timescale was significant, even when starting from all-infected. This network enables us to illustrate the numerical measurement of our quantifiers in a scenario where the QSD has less practical relevance, i.e. where endemic quasi-stationary behaviour and dichotomised persistence are not `recognisable' phenomena. In this case, we can compute $P_{T,\Gamma}^i(\mbox{quasi-invasion})$ ($=P_{T,\Gamma}^i(\mbox{quasi-prevalence})$) by directly measuring $P_{T,\Gamma}(\xi_t^i \cap V \neq \emptyset)/ P_{T,\Gamma}( \xi_t^V \cap V \neq \emptyset)$ at increasing time points and then estimating its convergent value. Thus,  for two different global transmission parameters (0.8, 0.5), and two different initial states (all-infected, one infected), 1 million simulations were allowed to run up to some specific point in simulated time (the global recovery parameter was always set to 1). For each simulation, the time at which extinction occurred was recorded so that the probability of survival up to increasing points in time could be measured.

 In figure~\ref{F5}a, our quantifier is able to capture an important feature of the model, i.e.  the value at which $P_{T,\Gamma}(\xi_t^i \cap V \neq \emptyset)$ $(=P_{T,\Gamma}(\xi_t^V \cap i \neq \emptyset))$ plateaus before its inevitable decay to zero. Figure~\ref{F5}b gives an example of a scenario where, although our quasi-invasion and quasi-prevalence quantifiers are clearly defined, their practical relevance is less obvious. This is because the transmission parameter was sufficiently low such that early extinction was the dominant behaviour. 

\subsection*{Computational Efficiency in the Measurement of Invasion Probability and Endemic Prevalence - A New Perspective}

 Through duality, we can approximate $P_{T,\Gamma}^A(\mbox{quasi-prevalence})$ by measuring $P_{T^T,\Gamma}(\xi_t^A \cap V \neq \emptyset)$ at increasing points in time (as in figure~\ref{F5}) in order to estimate the value at which it may initially plateau. This could, in certain circumstances, be much more efficient than trying to establish global quasi-stationary behaviour and then computing the proportion of time for which the infection is present in $A$.  Conversely, if we wish to approximate $P_{T,\Gamma}^i(\mbox{quasi-invasion})$, for all $i \in V$, it may be more computationally efficient to first establish global quasi-stationary behaviour in the transposed network and then measure the proportion of time each individual spends infected.

\section*{Discussion}

By considering the unique QSD associated with Markovian SIS dynamics on finite strongly connected networks, along with its implications under duality, we have provided meaningful mathematical definitions for both endemic prevalence (quasi-prevalence) and invasion probability (quasi-invasion). Utilising these definitions, we have also provided a general statement of the exact relationship between invasion probability and endemic prevalence at the individual and population level, for any finite undirected network of arbitrary heterogeneity (including undirected networks with weighted links and individual-specific recovery parameters). The relationship also has implications in the context of directed networks.

We note that for two specific homogeneous networks (infinite square lattice and infinite `great circle'), invasion probability (in these cases, the probability of indefinite persistence) from a single initial infected has been shown to be equal to the fraction of the population infected in the upper invariant measure \cite{Grimmett}, \cite{Neal}. Furthermore, the relationship between the probability of long-term persistence and quasi-stationary distributions has previously been investigated (see Chaterjee and Durrett \cite{Chaterjee} and, for the related concept of `metastability', see Schonmann  \cite{Schonmann} and Simonis  \cite{Simonis}). However, although the prevalence-invasion relationship follows easily from a combination of the QSD and duality, to our knowledge this is the first general statement of this exact relationship for any finite strongly connected network. We have thus related two fundamental epidemiological quantifiers in systems where they cannot usually be calculated analytically due to complexity.

It is generally easier to collect empirical data on endemic prevalence rather than directly on invasion risk. In the case of undirected networks, prevalence data can thus be utilised to inform invasion risk. This method echoes Anderson and May's \cite{Anderson2} estimation of the `basic reproductive ratio' of measles from the total number of susceptible individuals in England and Wales (using data from Fine and Clarkson \cite{Fine}). When other infectious agents exhibit qualitatively similar behaviour on the same undirected network, we can expect that the individuals carrying the greatest level of endemic infection are also those most likely to initiate new successful invasions. This lends support to the targeting of high-risk individuals in these systems as an effective strategy for the mitigation and control of emerging epidemics.

\section*{Acknowledgments}

The authors thank Megan Selbach-Allen for discussions, Jane Rees for comments on the final manuscript, Art Jonkers for assistance in producing the example network and Ian Smith for assistance with high throughput computing. We thank two anonymous reviewers for comments which enhanced the clarity of the manuscript.

\end{document}